\documentclass[epj]{svjour}
\usepackage{graphics}
\usepackage{amssymb}
\begin{document}
\title{Grand Canonical simulations of string tension in elastic surface model}
\author{Hiroshi Koibuchi 
}                     
%
%
\institute{Department of Mechanical and Systems Engineering, \\
  Ibaraki College of Technology \\
  Nakane 866, Hitachinaka, Ibaraki 312-8508, Japan }
%
%
\abstract{
We report a numerical evidence that the string tension $\sigma$ can be viewed as an order parameter of the phase transition, which separates the smooth phase from the crumpled one, in the fluid surface model of Helfrich and Polyakov-Kleinert. The model is defined on spherical surfaces with two fixed vertices of distance $L$. The string tension $\sigma$ is calculated by regarding the surface as a string connecting the two points. We find that the phase transition strengthens as $L$ is increased, and that $\sigma$ vanishes in the crumpled phase and non-vanishes in the smooth phase. }
\PACS{
      {64.60.-i}{General studies of phase transitions} \and
      {68.60.-p}{Physical properties of thin films, nonelectronic} \and
      {87.16.Dg}{Membranes, bilayers, and vesicles}
} 
\maketitle
%
\section{Introduction}
A considerable number of studies have been conducted on the phase structure of Helfrich and Polyakov-Kleinert model of membranes \cite{POLYAKOV-NPB-1986,Kleinert-PLB-1986,HELFRICH-NF-1973,DavidGuitter-EPL-1988,Peliti-Leibler-PRL-1985,LEIBLER-SMMS-2004,BKS-PLA-2000,BK-PRB-2001,Kleinert-EPJB-1999,NELSON-SMMS-2004,KANTOR-SMMS-2004,BOWICK-TRAVESSET-PREP-2001,WIESE-PTCP19-2000}. The triangulated surface model can be classified into two groups \cite{David-TDQG-1989,WHEATER-JP-1994}. One is the model of fixed connectivity surfaces, and the other the dynamical connectivity surfaces, which are called fluid surfaces. Both kinds of surfaces become smooth (crumpled) at infinite (zero) bending rigidity $b$. The model on fixed connectivity surfaces has been considered to undergo a finite-$b$ transition between the smooth phase and the crumpled phase. A lot of numerical studies including those on fluid surfaces so far support this fact \cite{BOWICK-SMMS-2004,WHEATER-NPB-1996,BCFTA-1996-1997,AMBJORN-NPB-1993,CATTERALL-NPB-SUPL-1991,BCHGM-NPB-1993,ABGFHHE-PLB-1993,KANTOR-NELSON-PRA-1987,GOMPPER-KROLL-PRE-1995,KOIB-PLA-2002,KOIB-PLA-2003-2,BCTT-EPJE-2001,KOIB-PLA-2004}. However, there seems to be no established understanding of phase transitions in the fluid surface model.

Ambjorn et. al. have studied a mass gap and a string tension of the fluid model \cite{AMBJORN-NPB-1993}. It was reported in \cite{AMBJORN-NPB-1993} that the mass gap and the string tension vanish at the critical point of the phase transition, which has been considered not to be characterized by a divergence of the specific heat. The mass gap was extracted by assuming the spherical surface as an oblong one-dimensional string with fixed end points separated by a distance $L$. The string tension was also computed by assuming a surface as a sheet of area $A$ with fixed boundary, and the same results as those of the mass gap were obtained. They used the canonical Monte Carlo simulations, which are equivalent with the grand canonical ones.

Recent numerical simulations on the fluid surface model suggested that the phase transition is characterized by a divergence of the specific heat, although the parameter $\alpha$, the coefficient of the co-ordination dependent term, was assumed to have arbitrary values \cite{KOIB-PLA-2002,KOIB-PLA-2003-2}. Therefore, it is interesting to see whether the string tension vanishes or not at the critical point of the transition of the model with arbitrary $\alpha$. The notation {\it string tension} in this paper corresponds not to the string tension in  \cite{AMBJORN-NPB-1993} but to the mass gap in  \cite{AMBJORN-NPB-1993}; we use {\it string tension} in place of the mass gap and denote it by $\sigma$  hence force.  

From the simulation studies on the fluid model, we obtained a numerical evidence that the string tension $\sigma$ vanishes in the crumpled phase and non-vanishes in the smooth phase \cite{KOIB-PLA-2004}. The result presented in \cite{KOIB-PLA-2004} implies that $\sigma$ can be considered as an order parameter of the phase transition. 

The string tension is considered to be a key to understand the phase structure of the fluid surfaces. Therefore, we show in this paper our simulation data including those presented in \cite{KOIB-PLA-2004} in order to have an insight into further investigations on the phase structure of fluid surfaces. 

We comment on why the result of non-vanishing string tension could be a relevant one. It is possible to consider that the non-vanishing string tension is connected to two interesting problems. The first is the problem of quark confinement, which is a problem in high-energy physics. The linear potential $V(L)\sim L$ assumed between quark and anti-quark separated by distance $L$ gives a finite string tension, which is compatible with our result of non-vanishing string tension. 

The second is a conversion of external forces into an internal energy and vice versa in real physical membranes, and is a rather practical problem. We can consider that the transition depends on the temperature: the surface becomes crumpled at $T\!>\!T_c$ and smooth at $T\!<\!T_c$, where $T_c$ is the transition temperature. Then the surface is picked up in two points and extended to sufficiently large at $T\!>\!T_c$ in the beginning; this can be done with zero external force because of the zero string tension. Then, lowering the temperature to $T\!<\!T_c$, we have a finite string tension between the two points on the surface. This is a conversion of the internal energy $S$ into an external force. Conversely, an external force enlarging the surface in the smooth phase can be accumulated as an internal energy. This is also easy to understand because the free energy of a string can be written as (tension)$\times$(length).  If the model in this paper represents properties in some real membranes, our result implies a possibility of such conversions.  

\section{Model}
A sphere in ${\bf R}^3$ is discretized with piecewise linear triangles. Every vertex is connected to its neighboring vertices by bonds, which are the edges of triangles. Two vertices are fixed as the boundary points separated by a distance $L$.

The Gaussian energy $S_1$ and the bending energy $S_2$ are defined by
\begin{equation}
\label{S13-DISC}
S_1=\sum_{(ij)} \left(X_i-X_j\right)^2, \; S_2= \sum_i \left(1-\cos \theta_i\right), 
\end{equation}
where $\sum_{(ij)}$ is the sum over bonds $(ij)$, $\theta_i$ in $S_2$ is the angle between two triangles sharing the edge $i$, and $X_i (\in {\bf R}^3)$  the position of the vertex $i$.

The partition function $Z$ is defined by 
\begin{eqnarray}
 \label{Z-FLUID}
Z(b,\mu,\alpha; L) = \sum_N \sum_{T} \int \prod _{i=1}^N dX_i \exp\left[-S(X,{T},N)\right],\qquad \\
S(X,{T},N)=S_1 + b S_2 -\mu N - \alpha \sum_i \log q_i, \qquad\quad \nonumber
\end{eqnarray}
where $\sum_{T}$ denotes the sum over all possible triangulations ${T}$, and $N$ the total number of vertices. It should be noted that the chemical potential term $-\mu N$ and the co-ordination dependent term  $-\alpha \sum_i \log q_i$ are included in the Hamiltonian. The expression $S(X,{T},N)$  shows that $S$ explicitly depends on the variables $X$, ${T}$ and $N$. The coefficient $b$ is the bending rigidity, and $\mu$ the chemical potential. $Z$ depends on $b$, $\mu$, $\alpha$, and $L$. The surfaces are allowed to self-intersect and hence phantom.

We consider that the phase structure of the model depends on the choice of the integration measure $\prod_i q_i^{\alpha} dX_i$, where $q_i$ is the co-ordination number of the vertex $i$ \cite{KOIB-PLA-2002,KOIB-PLA-2003-2}. The co-ordination dependent term in Eq. (\ref{Z-FLUID}) comes from this integration measure, because $\prod_i q_i^{\alpha} dX_i$ can also be written as 
$\prod_i dX_i \exp (\alpha \sum_i \log q_i)$. This $\alpha$ is believed to be $2\alpha\!=\!3$ \cite{DAVID-NPB-1985,ADF-NPB-1985,FN-NPB-1993}, and hence it is unclear whether $\alpha$ can take arbitrary value. On the other hand $q_i^{\alpha}$ is considered as a volume weight of the vertex $i$ in the integration $dX_i$. Thus it is possible to extend $\alpha$ to continuous numbers by assuming that the weight takes a suitable value. Therefore, it is interesting to see the dependence of $\sigma$ on the phase transitions which can be controlled by the parameter $\alpha$. We note that the continuous $\alpha$ assumed in our model does not influence $2\alpha\!=\!3$ in the model of \cite{DAVID-NPB-1985,ADF-NPB-1985,FN-NPB-1993}.

Note also that the constant term $-\alpha \sum_i \log 6$ can be included in $S$ of Eq. (\ref{Z-FLUID}), because $S$ can be written as $S \!=\!S_1 + b S_2 -\mu^\prime N - \alpha \sum_i \log (q_i/6)$, where $\mu\!=\!\mu^\prime\!-\!\alpha \log 6$. As a consequence, the total number $N$ of vertices depends on $\mu$ and $\alpha$ in the grand canonical simulations using $S$ that does not include the constant term. If the simulations were done by using $S$ that includes the constant term, the results must be equivalent with those without the constant term because of the relation between $\mu$ and $\mu^\prime$ described above.

Let us comment on a relation between the value of $\alpha$ and that of $q^{\rm max}$ the maximum co-ordination number, and consider why the phase transition is sensitive to the value of $\alpha$. The reason why the phase transition is strengthened at negative $\alpha$ is that the co-ordination dependent term $-\alpha \sum_i \log q_i$ crumples the surface when $\alpha\!<\!0$ and competes with the bending energy term $bS_2$ smoothing the surface. On the contrary, the term $-\alpha \sum_i \log q_i$ tends to make $q$ such that $q\!\simeq\!6$ when $\alpha\!> \!0$. Because of the fact that $\sum_i q_i$ is constant on triangulated surfaces due to the topological constraint, $\sum_i \log q_i$ becomes maximum on the surfaces of uniform co-ordination number $q$. As a consequence, negative $\alpha$ make the surface non-uniform in $q$. Therefore, when $\alpha$ becomes negative large, then $q^{\rm max}$ increases, and the surface becomes crumpled. While the bending energy $S_2$ makes the surface smooth, the co-ordination dependent term with negative $\alpha$ makes the surface crumpled. Thus two competitive forces co-exist when $\alpha$ is negative: one is from the bending energy and the other from the co-ordination dependent term. 

We expect 
\begin{equation}
\label{tension}
Z(b,\mu, \alpha; L)\sim \exp(-\sigma L)
\end{equation}
 in the limit $L\to \infty$ \cite{AMBJORN-NPB-1993}. Then, by using the scale invariance of the partition function, we have  \cite{WHEATER-JP-1994,AMBJORN-NPB-1993}
\begin{equation}
\label{string-tension}
\sigma = {2 \langle S_1\rangle - 3 \langle N\rangle \over L},
\end{equation}
where $\langle S_1\rangle$  and $\langle N\rangle$ are the mean values of $S_1$ and $N$.

We note that a surface enclosing two fixed vertices is not a one-dimensional string, because the perpendicular size of the surface increases with $N$. However $L$ is chosen to be sufficiently larger than the perpendicular size, so that Eq. (\ref{tension}) holds.

The specific heat, which is the fluctuation of $S_2$, is defined by $C_{S_2}\!=\!(b^2/\langle  N\rangle ) (\partial^2 \log Z / \partial b^2)$, and is calculated by using 
\begin{equation}
\label{Spec-Heat-S2}
C_{S_2} = {b^2\over \langle N\rangle} \langle \; \left( S_2 - \langle S_2 \rangle\right)^2 \; \rangle.
\end{equation}
The fluctuation of $N$ denoted by $C_N$ can also be given by

\begin{equation}
\label{Spec-Heat-N}
C_{N} = {\mu^2\over \langle N\rangle} \langle \; \left( N- \langle N\rangle\right)^2 \; \rangle.
\end{equation}
As we will see later, the phase transition of the model is characterized by the divergence of $C_{S_2}$ and that of $C_N$.
\section{Monte Carlo technique}\label{MC-Techniques}
$X$ is updated so that $X^\prime \!=\! X \!+\! \delta X$, where $\delta X$ takes a value randomly in a small sphere. The radius $\delta r$ of the small sphere is chosen to maintain about $50 \%$ acceptance $r_X$ for the $X$-update. The radius $\delta r$ is defined by using a constant number $\epsilon$ as an input parameter so that $\delta r \!=\! \epsilon\, \langle l\rangle$, where $\langle l\rangle$ is the mean value of bond length computed at every 250 MCS (Monte Carlo sweeps). It should be noted that $\delta r$ is almost fixed because $\langle l\rangle$ is constant and unchanged in the equilibrium configurations. 

${T}$ is updated by flipping a bond shared by two triangles. The bonds are labeled by sequential numbers and chosen randomly to be flipped. The rate of acceptance $r_{\it T}$ for the bond flip is uncontrollable, and the value of $r_{\it T}$ is about $30\% \leq r_{\it T} \leq 40 \%$. $N$-trials for the updates of $X$ and $N$-trials for ${T}$ are done consecutively, and these make one MCS. 

$N$ is updated by both adsorption and desorption. In the desorption, a vertex is randomly chosen, and then a bond that is connected to the vertex is randomly chosen so that the two vertices at the ends of the bond unite and change to a new vertex. In the adsorption, a triangle is randomly chosen in the same way that a bond is chosen in the desorption, and a new vertex is added to the center of the triangle. As a consequence, the Euler number (=2) of the surface remains unchanged in the adsorption and the desorption. The acceptance rate $r_N$ is uncontrollable as $r_{T}$ is, and the value of $r_N$ is about $55\% \leq r_N \leq 65\%$ in our MC. 

In the adsorption of a vertex, the corresponding change of the total energy ${\it \Delta}S\!=\!S({\rm new})\!-\!S({\rm old}) $ is calculated. The adsorption is then accepted with the probability  \\
${\rm Min}[1,\exp\left(-{\it \Delta}S\right)/(N+1) ]$. In the desorption, \\
${\it \Delta}S\!=\!S({\rm new})\!-\!S({\rm old}) $ is calculated by assuming that one vertex is removed. The desorption is then accepted with the probability ${\rm Min}\left[1,N \exp\left(-{\it \Delta}S\right)\right]$. The adsorption and the desorption are tried alternately at every 5-MCS.

 We use surfaces of size $N\!\simeq\!500$, $N\!\simeq\!1000$, and $N\!\simeq\!1500$. The size $N$ depends on both $\mu$ and $\alpha$ which is fixed to three values: $\alpha\!=\!5.5$, $\alpha\!=\!0$, and $\alpha\!=\!-5.5$. The reason for choosing these three values of $\alpha$, the phase transition of the fluid surfaces is sensitive to $\alpha$ as noted in the previous section. The values of $\mu$ are chosen so that $N\!\simeq\!500$, $N\!\simeq\!1000$, and $N\!\simeq\!1500$ for each $\alpha$. The diameter $L_0(N)$ of the spheres at the start of MC simulations is fixed so that $\sum_i l_i^2\!\simeq\!3N/2$, where $l_i$ is the length of the bond $i$. As a consequence, $L_0(N)$ becomes
\begin{equation}
\label{L-scale}
L_0(N) \propto  \sqrt{N}.
\end{equation}
 We use three kinds of distance $L$ of the boundary points for each $L_0(N)$ such that $L\!=\!1.5L_0(N)$, $L\!=\!2L_0$,  and $L\!=\!3L_0(N)$. The distance is increased from $L_0(N)$ to $L$ in the first $5\!\times\! 10^6$ MCS.  

It should be noted that $L\!=\!1.5L_0(N)$, $L\!=\!2L_0$, and $L\!=\!3L_0(N)$ become $\infty$ in the thermodynamic limit $N\!\to\!\infty$ because of Eq.(\ref{L-scale}). Therefore, $\sigma$ defined by Eq.(\ref{tension}) can be extracted from these values of $L$ at sufficiently large $N$. Thus, the length $L$ in this paper depends on $N$ and hence does not strictly correspond to the one in \cite{AMBJORN-NPB-1993}. In fact, the value of $L$ in \cite{AMBJORN-NPB-1993} is chosen so that $t\!=\!L/N$ changes for a given $N$. However, as we will see, the scaling property of physical quantities, such as the dependence of $\sigma$ on $N$, obtained in this paper is compatible with those of $\sigma$ on $t$ in \cite{AMBJORN-NPB-1993}. 
\section{Results}\label{Results}
\begin{figure}[htb]
\center
\resizebox{0.45\textwidth}{!}{%
  \includegraphics{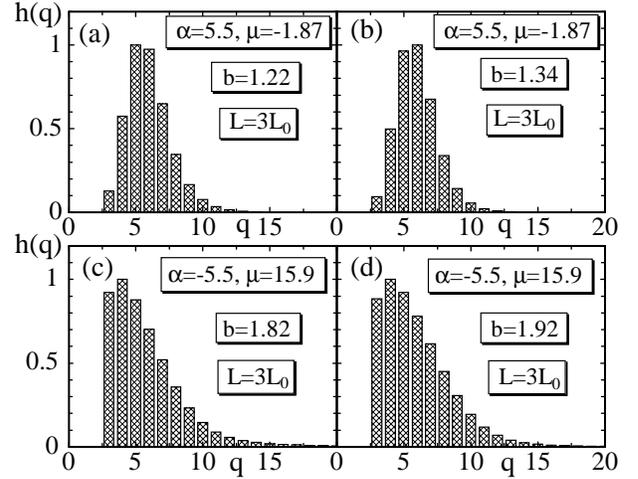}
}
 \caption{The normalized histogram $h(q)$ of the co-ordination number $q$ at (a) $\alpha=-5.5, \mu=-1.87$, $b=1.22$, (b) $\alpha=-5.5, \mu=-1.87$, $b=1.34$, (c) $\alpha=5.5, \mu=15.9$, $b=1.82$, 
and (d) $\alpha=5.5, \mu=15.9$, $b=1.92$. The histograms were obtained at the final $2\times 10^7$ MCS. }
\label{fig-1}
\end{figure}
\begin{figure}[htb]
\center
\resizebox{0.45\textwidth}{!}{%
  \includegraphics{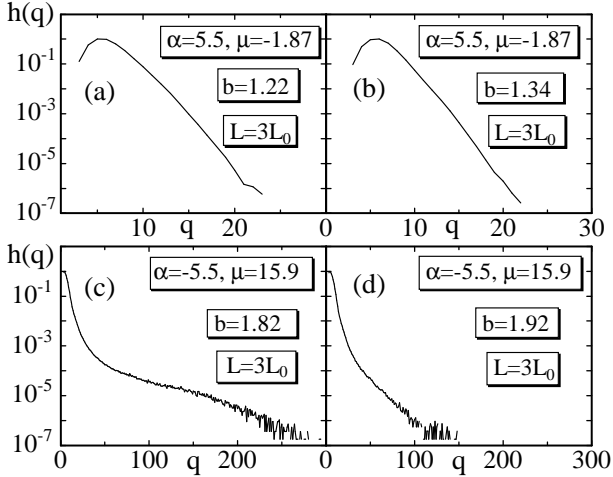}
}
 \caption{Plots of $\log h(q)$ against $q$ at (a) $\alpha=-5.5, \mu=-1.87$, $b=1.22$, (b) $\alpha=-5.5, \mu=-1.87$, $b=1.34$, (c) $\alpha=5.5, \mu=15.9$, $b=1.82$, and (d) $\alpha=5.5, \mu=15.9$, $b=1.92$, where $h(q)$ are those shown in Figs. \ref{fig-1}(a)--\ref{fig-1}(d). }
\label{fig-2}
\end{figure}
First, we show in Figs. \ref{fig-1}(a)--\ref{fig-1}(d) a normalized histogram $h(q)$ of the co-ordination number $q$, which is obtained during the final $2\times 10^7$ MCS at (a) $\alpha=-5.5, \mu=-1.87$, $b=1.22$, (b) $\alpha=-5.5, \mu=-1.87$, $b=1.34$, (c) $\alpha=5.5, \mu=15.9$, $b=1.82$, 
and (d) $\alpha=5.5, \mu=15.9$, $b=1.92$. The distance between the two vertices is $L=3L_0$, and the total number of vertices becomes $N\simeq 1500$ in those cases.  The surface becomes crumpled in (a) and smooth in (b), and there is no phase transition between these phases, as we will see below. We note also that the surface becomes crumpled in (c) and smooth in (d), and there is a first-order transition between these phases on the contrary. We see that the histograms shown in (a) and (b) are clearly different from those in (c) and (d). 

In order to show the difference more clearly, we plot $\log h(q)$ against $q$ in Figs. \ref{fig-2}(a)--\ref{fig-2}(d). We can see no co-ordination number of $q\geq 24$ in  Figs. \ref{fig-2}(a) and \ref{fig-2}(b). To the contrary, the curves in Figs. \ref{fig-2}(c) and \ref{fig-2}(d) indicate that there exist co-ordination numbers of $q\geq 200$ and $q\geq 100$ respectively. The curves of $h(q)$ in Figs. \ref{fig-2}(c) and \ref{fig-2}(d) indicate that configurations of large co-ordination numbers play some non-trivial role in the phase transition of fluid surfaces.   

\begin{figure}[htb]
\center
\resizebox{0.495\textwidth}{!}{%
  \includegraphics{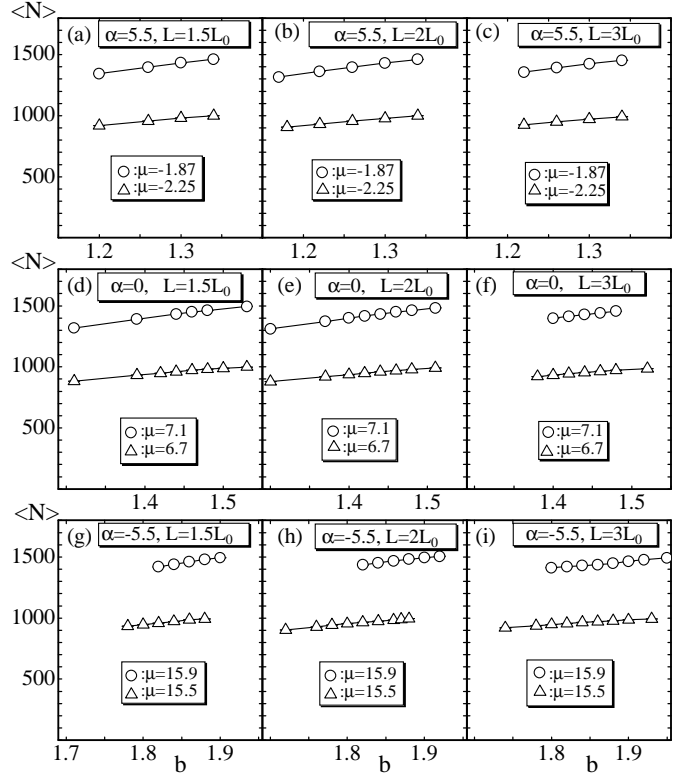}
}
 \caption{The average vertex number $\langle N \rangle$ against $b$ at (a) $\alpha\!=\!5.5$, $L\!=\!1.5L_0$, (b) $\alpha\!=\!5.5$, $L\!=\!2L_0$, (c) $\alpha\!=\!5.5$, $L\!=\!3L_0$, (d) $\alpha\!=\!0$, $L\!=\!1.5L_0$, (e) $\alpha\!=\!0$, $L\!=\!2L_0$, (f) $\alpha\!=\!0$, $L\!=\!3L_0$, (g) $\alpha\!=\!-5.5$, $L\!=\!1.5L_0$, (h) $\alpha\!=\!-5.5$, $L\!=\!2L_0$, and (i) $\alpha\!=\!-5.5$, $L\!=\!3L_0$. The symbols $\bigtriangleup $, and  $\bigcirc$ correspond to those obtained on surfaces of size $N\!\simeq\!1000$, and $N\!\simeq\!1500$ respectively. }
\label{fig-3}
\end{figure}
The average vertex number $\langle N \rangle$ are plotted in Figs. \ref{fig-3}(a)--\ref{fig-3}(i):  $\langle N \rangle$ in Figs. \ref{fig-3}(a), \ref{fig-3}(b), and \ref{fig-3}(c) are respectively obtained at $\alpha\!=\!5.5$, $L\!=\!1.5L_0$; $\alpha\!=\!5.5$, $L\!=\!2L_0$; and $\alpha\!=\!5.5$, $L\!=\!3L_0$. $\langle N \rangle$ in Figs. \ref{fig-3}(d), \ref{fig-3}(e), and \ref{fig-3}(f) are those at $\alpha\!=\!0$, $L\!=\!1.5L_0$; $\alpha\!=\!0$, $L\!=\!2L_0$; and $\alpha\!=\!0$, $L\!=\!3L_0$. $\langle N \rangle$ in Figs. \ref{fig-3}(g), \ref{fig-3}(h), and \ref{fig-3}(i) are those at $\alpha\!=\!-5.5$, $L\!=\!1.5L_0$; $\alpha\!=\!-5.5$, $L\!=\!2L_0$; and $\alpha\!=\!-5.5$, $L\!=\!3L_0$. The symbols $\bigtriangleup $, and  $\bigcirc$ in the figures correspond to those obtained on surfaces of size $N\!\simeq\!1000$, and $N\!\simeq\!1500$ respectively.      

We find from Figs. \ref{fig-3}(a)--\ref{fig-3}(i) that $\langle N \rangle$ is weakly dependent on $b$ and almost independent of $L$ with fixed $\mu$ and $\alpha$. The fluctuation $C_{N}$ of $\langle N \rangle$ can change against $b$ due to this dependence of $\langle N \rangle$ on $b$, and will be presented below.

\begin{figure}[htb]
\center
\resizebox{0.495\textwidth}{!}{%
  \includegraphics{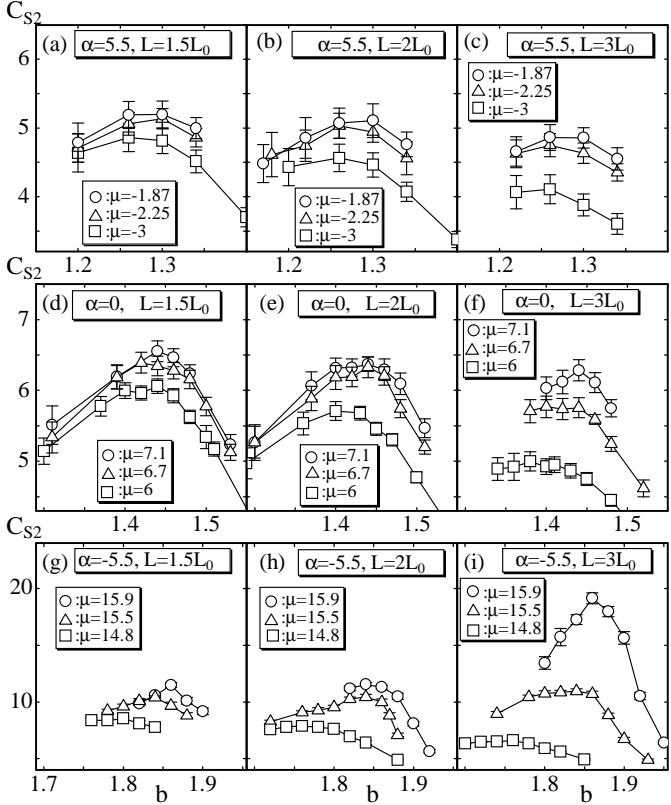}
}
 \caption{The specific heat $C_{S_2}$ against $b$ at (a) $\alpha\!=\!5.5$, $L\!=\!1.5L_0$, (b) $\alpha\!=\!5.5$, $L\!=\!2L_0$, (c) $\alpha\!=\!5.5$, $L\!=\!3L_0$, (d) $\alpha\!=\!0$, $L\!=\!1.5L_0$, (e) $\alpha\!=\!0$, $L\!=\!2L_0$, (f) $\alpha\!=\!0$, $L\!=\!3L_0$, (g) $\alpha\!=\!-5.5$, $L\!=\!1.5L_0$, (h) $\alpha\!=\!-5.5$, $L\!=\!2L_0$, and (i) $\alpha\!=\!-5.5$, $L\!=\!3L_0$. The symbols $\square $, $\bigtriangleup $, and  $\bigcirc$ correspond to those obtained on surfaces of size $N\!\simeq\!500$, $N\!\simeq\!1000$, and $N\!\simeq\!1500$ respectively. }
\label{fig-4}
\end{figure}
The specific heat $C_{S_2}$ defined by Eq. (\ref{Spec-Heat-S2}) are plotted in Figs. \ref{fig-4}(a)--\ref{fig-4}(i) and are respectively obtained at the same conditions for $\langle N \rangle$ shown in  Figs. \ref{fig-3}(a)--\ref{fig-3}(i).     
   
$C_{S_2}$ at $\alpha\!=\!5.5$ shown in Figs. \ref{fig-4}(a), \ref{fig-4}(b), and \ref{fig-4}(c) have peaks at intermediate $b$, however, the growth of peaks with increasing $N$ is almost invisible. On the contrary, we clearly see the growing of the peaks of $C_{S_2}$ at $\alpha\!=\!0, L\!=\!3L_0$ in Fig. \ref{fig-4}(f), and at $\alpha\!=\!-5.5$ in  Figs. \ref{fig-4}(g), \ref{fig-4}(h), and \ref{fig-4}(i).  These indicate that the phase transition strengthens not only with decreasing $\alpha$ but also with increasing $L$ at least in the region $-5.5 \!\leq\! \alpha \!\leq\! 0$.   

We comment on the total number of MCS and on the thermalization MCS. The convergence speed slows down when $\alpha$ decreases, because the maximum co-ordination number $q^{\rm max}$ increases with decreasing $\alpha$. $9.6\!\times\! 10^8$ MCS were done at $\alpha\!=\!-5.5$, $L\!=\!3L_0$, $b\!=\!1.86$, where $C_{S_2}$ has the peak; $7.6\!\times\! 10^8$ MCS at $\alpha\!=\!-5.5$, $L\!=\!2L_0$, $b\!=\!1.86$;  and $4\!\times\! 10^8$ MCS at $\alpha\!=\!-5.5$, $L\!=\!1.5L_0$, $b\!=\!1.86$. Relatively smaller number of MCS was done at $b$ that are distant from the transition point, and at $\alpha\!=\!0$, $\alpha\!=\!5.5$. The thermalization sweeps was about $1\!\times\! 10^7 \sim 3\!\times\! 10^7 $ on surfaces of $N\!\simeq\! 1500$ at $\alpha\!=\!-5.5$. Relatively smaller MCS for the thermalization were done in other cases. 

\begin{figure}[htb]
\center
\resizebox{0.495\textwidth}{!}{%
  \includegraphics{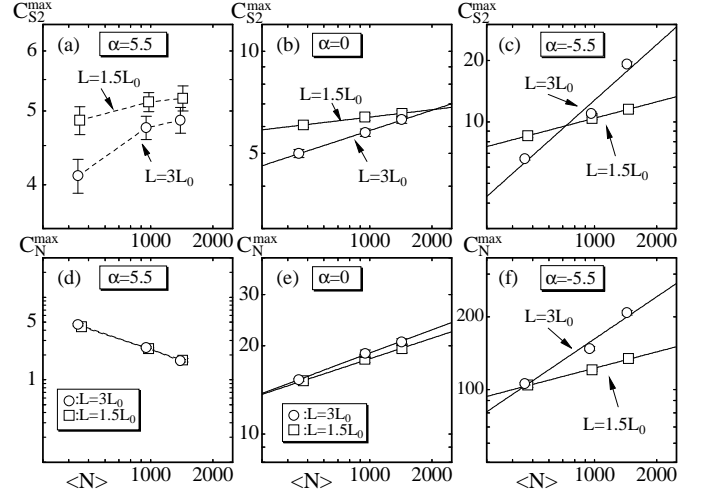}
}
 \caption{Log-log plots of the peak values $C_{S_2}^{\rm max}$ of the specific heat $C_{S_2}$ against $\langle N\rangle$ obtained at (a) $\alpha\!=\!5.5$,  (b) $\alpha\!=\!0$, and (c) $\alpha\!=\!-5.5$, and log-log plots of the peak values $C_N^{\rm max}$ of $C_N$ against $\langle N\rangle$ at (d) $\alpha\!=\!5.5$,  (e) $\alpha\!=\!0$, and (f) $\alpha\!=\!-5.5$.} 
\label{fig-5}
\end{figure}
In order to see the scaling property of the peak values $C_{S_2}^{\rm max}$, we plot $C_{S_2}^{\rm max}$ against $N$ in log-log scales in Figs. \ref{fig-5}(a), \ref{fig-5}(b), and \ref{fig-5}(c) respectively obtained at $\alpha\!=\!5.5$, $\alpha\!=\!0$, and $\alpha\!=\!-5.5$. We find that $C_{S_2}^{\rm max}$ at $\alpha\!=\!5.5$ in Fig. \ref{fig-5}(a) saturate as $N$ increases. On the contrary, $C_{S_2}^{\rm max}$ at $\alpha\!=\!0$ in Fig. \ref{fig-5}(b) and those at $\alpha\!=\!-5.5$ in Fig. \ref{fig-5}(c) clearly scale according to
\begin{equation}
\label{C-scale-prop}
C_{S_2}^{\rm max} \propto  N^\nu.
\end{equation}
From the slope of the straight lines in Figs. \ref{fig-5}(b) and \ref{fig-5}(c), we have
\begin{eqnarray}
\label{scaling-exponent-1}
&&\nu = 0.027\pm0.025  \quad\left[\alpha\!=\!0, \; L\!=\!1.5L_0\right], \nonumber \\
&&\nu = 0.199\pm0.032  \quad \left[\alpha\!=\!0,\; L\!=\!3L_0\right]
\end{eqnarray}
and 
\begin{eqnarray}
\label{scaling-exponent-2}
&&\nu=0.265\pm 0.025 \quad \left[\alpha\!=\!-5.5, \; L\!=\!1.5L_0\right], \nonumber \\
&&\nu=0.822\pm0.182 \quad \left[\alpha\!=\!-5.5,\; L\!=\!3L_0\right]. 
\end{eqnarray}
From the value $\nu\!=\!0.199\!\pm\!0.032$ at $\alpha\!=\!0, L\!=\!3L_0$ in Eq. (\ref{scaling-exponent-1}) and that $\nu\!=\!0.265\!\pm\!0.025$ at $\alpha\!=\!-5.5, L\!=\!1.5L_0$ in Eq. (\ref{scaling-exponent-2}), we understand that the surfaces undergo continuous transitions at those conditions.  Moreover, $\nu\!=\!0.822\!\pm\!0.182$ in Eq.(\ref{scaling-exponent-2}) indicates that the phase transition is of first order.

The peak values $C_N^{\rm max}$ of the specific heat $C_N$, which is the fluctuation of $N$ defined by Eq. (\ref{Spec-Heat-N}), is plotted in Figs. \ref{fig-5}(d), \ref{fig-5}(e), and \ref{fig-5}(f).  We find also from these figures of $C_N^{\rm max}$ that the phase transition occurs at $\alpha\!=\!0$ and $\alpha\!=\!-5.5$, and that there is no phase transition at $\alpha\!=\!5.5$. Thus, we confirm that the phase structure described by $C_N$ is compatible with that by $C_{S_2}$. 

\begin{figure}[hbt]
\center
\resizebox{0.495\textwidth}{!}{%
  \includegraphics{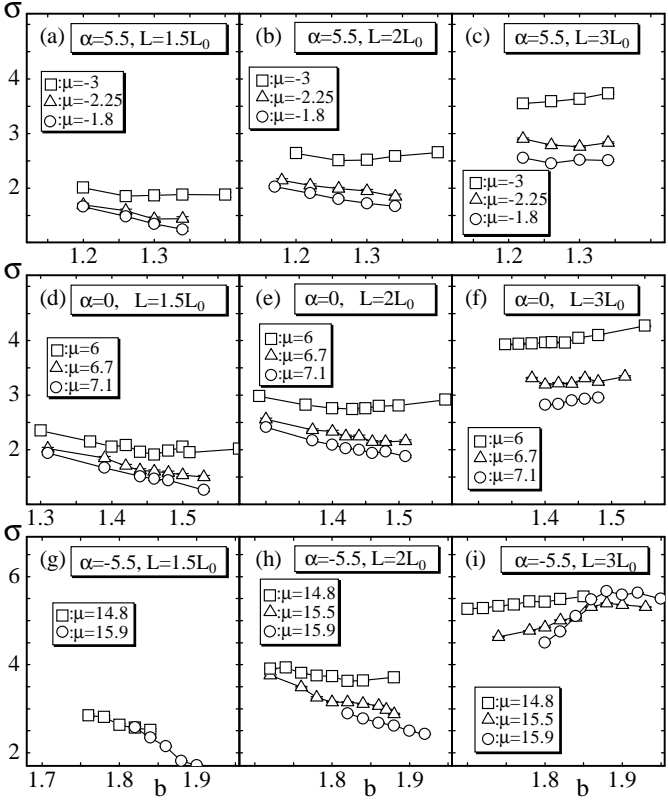}
}
\caption{String tension $\sigma$ against $b$ obtained at (a) $\alpha\!=\!5.5$, $L\!=\!1.5L_0$, (b) $\alpha\!=\!5.5$, $L\!=\!2L_0$, (c) $\alpha\!=\!5.5$, $L\!=\!3L_0$, (d) $\alpha\!=\!0$, $L\!=\!1.5L_0$, (e) $\alpha\!=\!0$, $L\!=\!2L_0$, (f) $\alpha\!=\!0$, $L\!=\!3L_0$, (g) $\alpha\!=\!-5.5$, $L\!=\!1.5L_0$, (h) $\alpha\!=\!-5.5$, $L\!=\!2L_0$, and (i) $\alpha\!=\!-5.5$, $L\!=\!3L_0$. The symbols $\square $, $\bigtriangleup $, and  $\bigcirc$ correspond to those obtained on surfaces of size $N\!\simeq\!500$, $N\!\simeq\!1000$, and $N\!\simeq\!1500$ respectively. }
\label{fig-6}
\end{figure}
Figures \ref{fig-6}(a)--\ref{fig-6}(i) are plots of the string tension $\sigma$ against $b$, which were obtained under the conditions that are exactly same as those in Figs. \ref{fig-4}(a)--\ref{fig-4}(i). The string tension $\sigma$ is calculated by Eq. (\ref{string-tension}). We see in Figs. \ref{fig-6}(a),  \ref{fig-6}(b), and \ref{fig-6}(c) that $\sigma$ increases with increasing $L$ and that $\sigma$ on surfaces of small $N$ is relatively larger than that of larger surfaces. It is also understood from Figs. \ref{fig-6}(a) and  \ref{fig-6}(b) that $\sigma$ decreases with increasing $b$ on larger surfaces. This indicates that $\sigma$ on smooth surfaces are larger than those on crumpled surfaces. These properties of $\sigma$ can be seen in those obtained at $\alpha\!=\!0$ in Figs. \ref{fig-6}(d) and \ref{fig-6}(e), and also seen in those  obtained at $\alpha\!=\!-5.5$ in Figs. \ref{fig-6}(g) and  \ref{fig-6}(h). On the contrary, we find from Fig. \ref{fig-6}(i) that $\sigma$ rapidly changes at the transition point when $N$ is increased. We already saw in Fig. \ref{fig-4}(i) that $b\!=\!1.86$ is the transition point of the surface of size $N\!\simeq \! 1500$ at $\alpha\!=\!-5.5$, $L\!=\!3L_0$. Therefore, we can see in Fig. \ref{fig-6}(i) that the string tension $\sigma$ vanishes in the crumpled phase and non-vanishes in the smooth phase.  

\begin{figure}[htb]
 \center
 \resizebox{0.495\textwidth}{!}{%
  \includegraphics{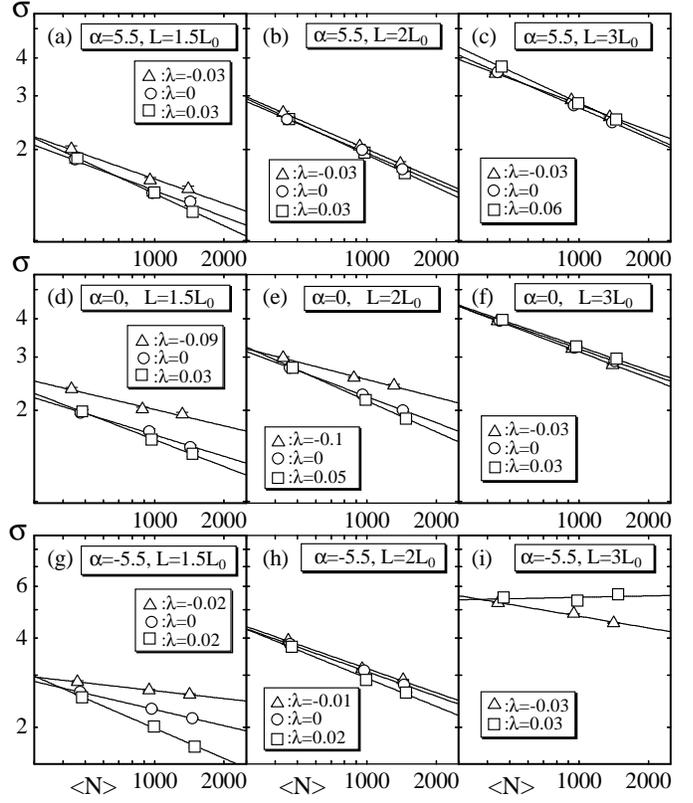}
}
\caption{Log-log plots of the string tension $\sigma$ against $\langle N\rangle$ obtained at  (a) $\alpha\!=\!5.5$, $L\!=\!1.5L_0$, (b) $\alpha\!=\!5.5$, $L\!=\!2L_0$, (c) $\alpha\!=\!5.5$, $L\!=\!3L_0$, (d) $\alpha\!=\!0$, $L\!=\!1.5L_0$, (e) $\alpha\!=\!0$, $L\!=\!2L_0$, (f) $\alpha\!=\!0$, $L\!=\!3L_0$, (g) $\alpha\!=\!-5.5$, $L\!=\!1.5L_0$, (h) $\alpha\!=\!-5.5$, $L\!=\!2L_0$, and (i) $\alpha\!=\!-5.5$, $L\!=\!3L_0$.}
\label{fig-7}
\end{figure}
In order to see the scaling property of $\sigma$, we introduce the reduced bending rigidity $\lambda$ such that
\begin{equation}
\label{reduced-b}
\lambda = {b \over b_c(\mu,\alpha)} - 1,
\end{equation}
where $b_c(\mu,\alpha)$ denotes the transition point where $C_{S_2}$ has the peak value shown in Figs. \ref{fig-4}(a)-- \ref{fig-4}(i). Then, the transition point $b_c(\mu,\alpha)$ is represented by $\lambda\!=\!0$, the smooth phase at $b\!>\!b_c(\mu,\alpha)$ by $\lambda\!>\!0$, and the crumpled phase at $b\!<\!b_c(\mu,\alpha)$ by $\lambda\!<\!0$. The reason why we introduce $\lambda$ of Eq. (\ref{reduced-b}) is because the transition point $b_c(\mu,\alpha)$ moves right as $N$ increases, as confirmed in Fig. \ref{fig-4}(i) for example.

Figures \ref{fig-7}(a)-- \ref{fig-7}(i) show log-log plots of $\sigma$ against $\langle N\rangle$ obtained at $\lambda\!<\!0$, $\lambda\!=\!0$, and $\lambda\!>\!0$. The straight lines in each figure denote the scaling property of $\sigma$ such as
\begin{equation}
\label{sigma-scale}
\sigma \propto N^{-\kappa} \qquad(\kappa \geq 0).
\end{equation}
We confirm from Fig. \ref{fig-7}(i) that $\sigma$ non-vanishes at $\lambda\!=\!0.03$ in the smooth phase, which was expected also from Fig. \ref{fig-6}(i) \cite{comment-1}. Moreover, we find from Figs. \ref{fig-7}(a)-- \ref{fig-7}(i), and  Eq. (\ref{kappa-results}) that almost all $\sigma$ satisfy $\sigma\!\to\!0(N\!\to\!\infty)$, which is the scaling property at the continuous transition in \cite{AMBJORN-NPB-1993}. Recalling that continuous transitions can be seen at $\alpha\!=\!0$ and $\alpha\!=\!-5.5$ in Figs. \ref{fig-5}(b) and \ref{fig-5}(c) [or \ref{fig-5}(e) and \ref{fig-5}(f)],  we understand that the scaling of $\sigma$ shown in Figs. \ref{fig-7}(a)-- \ref{fig-7}(i), except the non-vanishing $\sigma$, are compatible with that in \cite{AMBJORN-NPB-1993}. 
  
The exponent $\kappa$ in Eq. (\ref{sigma-scale}) can be obtained by a least squares fitting, and some of the results are as follows:
\begin{eqnarray}
\label{kappa-results}
&&\kappa = 0.268\pm0.006  \; \left[\alpha\!=\!0, L\!=\!3L_0, \lambda\!=\!0\right],
\\
&&\kappa = 0.126\pm0.019  \; \left[\alpha\!=\!-5.5, L\!=\!3L_0, \lambda\!=\!-0.03\right], \nonumber \\
&&\kappa = -0.017\pm0.041  \; \left[\alpha\!=\!-5.5, L\!=\!3L_0, \lambda\!=\!0.03\right].
\nonumber 
\end{eqnarray}
The first $\kappa$ in Eq. (\ref{kappa-results}) was obtained at a continuous transition point, and the second and the third were at the discontinuous transition. Although $\kappa \!=\! -0.017(41)$ in the last of Eq. (\ref{kappa-results}) appears to be ill-defined, we consider that it is compatible with the non-vanishing string tension.

It should be emphasized that the scaling of $\sigma$ in Eq. (\ref{sigma-scale}) is compatible with  $\sigma \!\propto\! (L/N)^{\delta} $ with $\delta \!>\!0$  in \cite{AMBJORN-NPB-1993},  since $L\!\propto\! L_0(N)\!\propto\! \sqrt{N}$ as described in Eq. (\ref{L-scale}). $L_0(N)$ is the diameter of the initial sphere for the MC simulations and chosen to $ L_0(N)\!\propto\! \sqrt{N}$ as already noted in Eq. (\ref{L-scale}). 

In fact, we note that $\delta\!=\!2\kappa$ and $\delta$ corresponds to $\nu/(1\!-\!\nu)$ in Ref. \cite{AMBJORN-NPB-1993}, where $\nu$ is about $\nu\!\simeq\!0.28$ in the crumpled phase and $\nu\!\simeq\! 0.42$ in the smooth phase close to the critical point. These $\nu$ corresponds to $\kappa\!\simeq\!0.19$ and  $\kappa\!\simeq\!0.36$ respectively. Thus, these values of $\kappa$ in Ref. \cite{AMBJORN-NPB-1993} are roughly consistent with the result $\kappa\!=\! 0.268(6)$ in Eq. (\ref{kappa-results}), obtained at a continuous transition point.  
\section{Summary and Conclusions}
We have studied the phase structure of the fluid surface model of Helfrich and Polyakov-Kleinert by grand canonical simulations on spherical surfaces with two fixed vertices of distance $L$. The model is defined by Hamiltonian $S$ containing the Gaussian term $S_1$, the bending energy term $S_2$, the co-ordination dependent term $S_3$, and the chemical potential term $-\mu N$: $S\!=\!S_1\!+\!bS_2\!-\!\alpha S_3\!-\!\mu N$. It is expected that the model undergoes a finite-$b$ transition between the smooth phase at $b\!\to\!\infty$ and the crumpled phase at $b\!\to\!0$. The phase transition was observed at $\alpha\!=\!0$ and $\alpha\!=\!-5.5$. The order of the transition changes from second to first at $\alpha\!=\!-5.5$  with sufficiently large $L$. The string tension $\sigma$ was obtained by regarding the surface as a string connecting the two vertices. It is remarkable that $\sigma$ becomes nonzero in the smooth phase separated by the discontinuous transition from the crumpled phase. Our results indicate that $\sigma$ can be viewed as an order parameter of the phase transition. It should be noted that our results are compatible with those in \cite{AMBJORN-NPB-1993}, because the obtained $\sigma$ in our study vanishes at the critical point of the continuous transition. 

As we have confirmed in this paper, configurations of large co-ordination number appear in certain cases and play some non-trivial role in the phase transition. Although we have no clear interpretation of a broad distribution of co-ordination number, it is possible that the existence of large co-ordination number is connected with some heterogeneous structure of fluid surfaces. 

The results presented in this paper are not conclusive. Some problems remain to be studied: Can we find  a finite string tension in the smooth phase separated by a second-order transition from the crumpled one? Can we find that the order of the transition remains unchanged on larger surfaces? Can we find a clear interpretation of a broad distribution of the co-ordination number in biological membranes? We consider that some points can be resolved by the grand canonical MC simulations on sufficiently large surfaces. We expect that the non-vanishing string tension can also be obtained by the canonical Monte Carlo simulations on fluid surfaces. Further numerical studies would clarify the phase structure of the fluid model of Helfrich and Polyakov-Kleinert. 

This work is supported in part by a Grant-in-Aid for Scientific Research, No. 15560160. H.K. thanks N.Kusano, A.Nidaira, and K.Suzuki for their invaluable help.

%

\end{document}